

\input harvmac

\overfullrule=0pt




\rightline{USC-91/HEP-B5}
\rightline{October 1991}

\bigskip\bigskip

\centerline  {\bf GENERALIZED DUALITY AND SINGULAR STRINGS }
\centerline  {    {\bf IN HIGHER DIMENSIONS }
              {\footnote {$^*$}
              {   Research supported in part by the
              U.S. Department  of Energy, under Grant No.
              DE-FG03-84ER-40168
               }   }
                                   }


\vskip 1.00 true cm

\centerline { I. BARS and K. SFETSOS}

\bigskip

\centerline {Physics Department}
\centerline {University of Southern California}
\centerline {Los Angeles, CA 90089-0484, USA}

\vskip 1.50 true cm
\centerline{ABSTRACT}

Deformations of gauged WZW actions are constructed for any pair $(G,H)$ by
taking different embeddings of the gauge group $H\subset G$ as it acts on the
left and right of the group element $g$. This leads to models that are dual to
each other, generalizing the axial/vector duality of the two dimensional black
hole manifold. The classical equations are completely solved for any pair
$(G,H)$ and in particular for the anti de Sitter string based on $SO(d-
1,2)/SO(d-1,1)$ for which the normal modes are determined. Duality is
demonstrated for models that have the same set of normal modes. Concentrating
on $SO(2,2)/SO(2,1)$, the metric and dilaton fields of the $d=3$ string
as well as some of the dual generalizations are obtained. They have curvature
singularities and represent new singular solutions of Einstein's general
relativity in three dimensions.

\vfill\eject


\newsec { Introduction}

In order to study the behaviour of string propagation and interaction in
curved space {\it and} time anti-de-Sitter (ADS) non-compact current algebra
coset models $SO(d-1,2)/SO(d-1,1)$ were introduced as exact conformal theories
 \ref\BN{I. Bars and D. Nemeschansky, Nucl. Phys. B348 (1991) 89.}.
Using algebraic arguments as well as the equivalence of cosets to gauged
Wess-Zumino-Witten models
 \ref\WZW{E. Witten, Nucl. Phys. B223 (1983) 422.
 \semi K. Bardakci, E. Rabinovici and B. Saering, Nucl. Phys. B301 (1988) 151.
 \semi H.J. Schnitzer, Nucl. Phys. B324 (1989) 412.
  \semi D. Karabali, Q-Han Park, H.J. Schnitzer and Z. Yang, Phys. Lett. B216
(1989) 307.
 \semi D. Karabali and H.J. Schnitzer, Nucl. Phys. B329 (1990) 649.
 \semi K. Gawedzki and A. Kupiainen, Nucl. Phys. B320 (1989) 625.},
it was argued in \BN\ that these theories contain only one
time dimension and $d-1$ space dimensions for $d=2,3,4,\cdots 26$ (for
the supersymmetric versions $d=2,3,\cdots,10$).
The original motivation for these models was to try to learn about the
quantum properties of strings in non-trivial space-times (see also
 \ref\SL{J. Balog, L. O'Raifeartaigh, P. Forg\'ac and A. Wipf, Nucl. Phys.
B325 (1989) 225.\semi  P.M.S. Petropoulos, Phys. Lett. B326 (1990) 151.
 \semi S. Hwang, Nucl. Phys. B354 (1991) 100. } )
by using the techniques of non-compact current algebra
 \ref\DLP{L. Dixon, J. Lykken, and M. Peskin, Nucl. Phys. B325 (1989) 329.}
 \ref\IBNC{I. Bars, Nucl. Phys. B334 (1990) 125.}.
With Witten's interpretation
 \ref\WIT{E. Witten, Phys. Rev. D44 (1991) 314.}
of the $d=2$ model $SO(2,1)/SO(1,1)$  as a string propagating in a black hole
background in two space-time dimensions, it has become apparent that these
models may be able to shed some new light on gravitational singularities
 \ref\WAD{G. Mandal, A.M. Sengupta and S.R. Wadia, `` Classical Solutions of
$2d$ String Theory'', IASSNS-HEP-91/14.}.
 It is also possible to give a cosmological interpretation to the two
dimensional solution
 \ref\IBCS{I. Bars, ``Curved Space-Time Strings and Black Holes",
USC-91/HEP-B4.}
 \ref\TV{A.A. Tseytlin and C. Vafa, ``Elements of String Cosmology",
 HUTP-91/A49 }.
These interpretations are obtained by an essentially classical analysis.
The exact $d=2,\ k=9/4$ quantum theory has been developed partially [1,10-15]
and not surprizingly it reveals new features.
 \nref\IBBH{I. Bars,``String Propagation on Black Holes'', USC-91/HEP-B3.}
 \nref\DVV{R. Dijkgraaf, H. Verlinde and E. Verlinde, ``String Propagation in a
Black Hole Background", PUPT-1252 (1991)}
 \nref\BER{ M. Bershadsky and D. Kutasov, ``Comment on Gauged WZW Theory",
 HUTP-91/A024.}
 \nref\SFE{K. Sfetsos, ``Degeneracy of String States in 2-d Black Hole and a
New Derivation of SU(1,1) Parafermion Characters", USC-91/HEP-S1.}
 \nref\DN{J. Distler and P. Nelson, ``New Discrete States of Strings Near a
Black Hole, UPR-0462T or PUPT-1262.}
 \nref\BK{ I.Bachas and E. Kiritsis, ``Beyond the Large N limit: Non-linear
$W_\infty$ Symmetry of the $SL(2,R)/U(1)$ Coset Model", LBL-31213.}

These developments motivate us to investigate the classical and quantum
properties of such models in higher dimensions. The list of all gauged WZW
models that yield a single time coordinate was completed \IBCS\ by considering
the hermitian symmetric spaces (HSS) with the group $G$ non-compact and $H'$
the maximally compact subgroup. In this case the subgroup always has a $U(1)$
and one may write $H'=H\times U(1)$. For the gauge subgroup in WZW model
take $H$. Then the coset string is given by $G/H$ and, because
of a neccesarily negative central extension, it describes a string with one
time coordinate (associated with the $U(1)$) and $dim(G/H')$ space
coordinates. An example is $SU(2,1)/SU(2)$. The full list is
$SU(n,m)/SU(n)\times SU(m)$, $SO(n,2)/SO(n)$, $SO^*(2n)/SU(n)$,
$Sp^*(2n)/SU(n)$, $E^*_6/SO(10)$, $E^*_7/E_6$. The stars on $SO^*,\ Sp^*,\
E^*$ imply that the real forms for these groups are taken such that the $G/H$
generators are purely non-compact.
Further generalizations include supersymmetric versions \BN\ \IBCS\ and direct
products of any of the previous cases with any conformal theory that describes
purely space components of strings, such that the total central charge is
$c=26$ ($c=15$ with supersymmetry). Using this later approach in
conjuction with the original 2d black hole, a 3-dimensional charged black
string
 \ref\HH{J. Horne and G. Horowitz, ``Exact Black String Solutions in Three
Dimensions", UCSBTH-91-39.}
and super black p-branes
 \ref\GS{S. Giddings and A. Strominger, `` Exact Fivebranes in Critical String
Theory ", UCSBTH-91-35.}
have been constructed.

In this paper it will be shown that any gauged WZW model described by a pair
$(G,H)$ has generalizations that correspond to different embeddings of the
gauge group that acts on the left or the right of the group element $g$.
Although our main motivation is to study the space-time strings described
above, the general remarks below apply to any pair $(G,H)$ for compact as
well as non-compact groups or (with appropriate modifications) supergroups.
The generalized action is

\eqn\naction{ \eqalign {
 &S=S_0(g)-{k\over 4\pi}\int_M d^2\sigma Tr(A_-\partial_+gg^{-1}- \tilde
A_+g^{-1}\partial_-g + A_-g\tilde A_+g^{-1}-A_-A_+)\cr
 &S_0={k\over 8\pi}\int_M d^2\sigma Tr(g^{-1}\partial_+g\ g^{-1}\partial_-g)
 -{k\over 24\pi}\int_B Tr(g^{-1}dg\ g^{-1}dg\ g^{-1}dg) }      }
where $g(\sigma^+,\sigma^-)$ is a group element and $A_{\pm}(\sigma^+,\sigma^-
) $ are the gauge fields associated with $H$. Furthermore, $\tilde A_{\pm}$ and
$A_{\pm}$ are related by

\eqn\aa{ A_\pm= t_a A^a_\pm , \qquad \tilde A_\pm = \tilde t_a A^a_\pm ,}
where $t_a,\tilde t_a$ are matrix representations of the Lie algebra of $H$
embedded in a representation of $G$
 \foot{ The sign of $k$ has been changed compared to the
compact case so that $k>0$ for $c=26$ and simultaneosly the signature of
space-time has the correct sign \IBNC\ \BN\ . Furthermore, in the path
integral that defines the quantum theory, the last cubic term in $S_0$ is
uniquely defined only when the third homotopy group of G vanishes or is the
set of integers, $\pi_3(G)=0 \ or\ Z$ (see Witten in ref.2). If $\pi_3=0$ then
$k$ can be any real number, while for $\pi_3=Z$ $k$ must be an integer. For a
non-compact group $G$ the maximal {\it compact} subgroup $H_{max}$ determines
$\pi_3(G)=\pi_3(H_{max})$. If $H_{max}$ has any non-Abelian part then
$\pi_3(H_{max})=Z$, and if it is purely Abelian $\pi_3(H_{max})=0$. Therefore
for all $d\ge 4$ ADS models and all HSS models one must take $k=integer$.
Accordingly, in order to obtain $c=26$ one may be forced to consider direct
product type models for some of the cases mentioned. One notable case, which
is not a direct product is the $d=4$ supersymmetric ADS model for which $c=15$
when $k=5$ \BN\ . Another one is the $d=5$ $SU(2,1)/SU(2)$ bosonic model which
gives $c=26$ for $k=4$ \IBCS\ .}.
The gauge invariance and conditions on $\tilde t_a$ will be given in section 3.

In section 2 we will solve generally the classical equations of the
standard gauged WZW model (with $\tilde A_\pm = A_\pm$) and in particular give
explicitly the full solution of the $d$ dimensional ADS model. In section 3
this solution is related to the classical solution of the
generalized action \naction . An interesting aspect of the 2d black hole is
that it has the property of duality that maps very different regions of the
manifold into each other (analog of $R\rightarrow 1/R$ duality)
 \ref\GIV{A. Giveon,  ``Target Space Duality and Stringy Black Holes", LBL-
30671} \IBBH\ \DVV\
 \ref\KIR{E. Kiritsis, ``Duality in Gauged WZW Models", LBL-30747.}
 \ref\TSE{A.A. Tseytlin, ``Duality and the Dilaton", JHU-TIPAC-91008.} \IBCS\ .
One wonders whether this is special to $SO(2,1)/SO(1,1)$ because of the
anomaly free choices of vector or axial gauging of the $SO(1,1)$. In section 3
it is shown how, under special circumstances, the new action generalizes the
duality concept to all gauged WZW models (compact or non-compact). This
duality is applicable even with non-Abelian gauge subgroups. After these
general observations, in section 4 we concentrate on the $d=3$ ADS string and
derive the classical gravitational metric and dilaton fields.
The dually related metrics are also discussed. These metrics are
both time and space dependent and have curvature singularities. The 3d
space-time manifolds described by these metrics provide  new and unfamiliar
singular solutions of Einstein's equations.


\newsec {Classical Solution}

The equations of motion that follow from \naction\ , when $\tilde A_\pm
=A_\pm$, are
\eqn\class{ (D_+gg^{-1})_{H}=(g^{-1}D_-g)_{H}=0,\qquad F_{+-}=0,\qquad
D_-(D_+gg^{-1})_{G/H}=0 , }
where $D_{\pm}g=\partial_{\pm}g-[A_{\pm},g]$ and $F_{+-}=\partial_+A_--
\partial_-A_+-[A_+,A_-]$. The subscripts $H,\ G/H$ imply that the matrices
that represent the quantities in parantheses have to be restricted to the
$H$-subspace or $G/H$-subspace.
$F_{+-}=0$ requires a pure gauge field $A_{\pm}=-\tilde
h^{-1}\partial_{\pm}\tilde h$ (locally). Choosing the gauge in which $\tilde
h=1$ one finds $A_{\pm}=0$. In this gauge a global $H$ symmetry remains and
\class\ become
\eqn\cla{ (\partial_+g g^{-1})_H=(g^{-1}\partial_-g)_H=0, \qquad
\partial_-(\partial_+gg^{-1}) =0 , }
where the $G/H$ subscript in the last equation is omitted since by using
the first equation the same result is achieved. The last equation is now
easily solved since it is the same as the ungauged WZW model
\eqn\sol{ g(\sigma^+,\sigma^-)= g_L(\sigma^+)g_R^{-1}(\sigma^-) }
where $g_L,g_R$ are group elements. From \cla\ they must satisfy the decoupled
equations
\eqn\dec{ (\partial_+g_Lg_L^{-1})_H=0,\qquad  (\partial_-g_Rg_R^{-1})_H=0 . }
To proceed further it is useful to parametrize the group elements as a product
$g_L=h_Lt_L$, $g_R=h_Rt_R$ where $h_L,h_R\in H$ and $t_L,t_R\in (G/H)$. Then
the equations take the form
\eqn\decoup{ h_L^{-1}\partial_+h_L + (\partial_+t_Lt_L^{-1})_H=0, \qquad
  h_R^{-1}\partial_-h_R + (\partial_-t_Rt_R^{-1})_H=0  }
where the $H$ projection needs to be applied only to the second term. These
equations are easily integrated
\eqn\hh{ h_L(\sigma^+)= P \big [e^{-\int^{\sigma^+} (\partial_+
t_Lt_L^{-1})_H}\big ], \qquad
 h_R(\sigma^-)= P \big [e^{-\int^{\sigma^-}
(\partial_- t_Rt_R^{-1})_H}\big ] ,  }
where $P$ stands for path ordering. Thus the full solution for a closed
string is obtained by considering all coset elements $t_L(\sigma^+),
t_R(\sigma^-)$ that are periodic in $\sigma^{\pm}$ (up to possible
isometries). The only restrictions on $t_L,t_R$ come from conformal
invariance in the form of vanishing stress tensor $T_L,T_R$ for left and right
movers. Since with \hh\ the subgroup currents are identically zero, the stress
tensor is constructed from only the coset currents
$(\partial_+g_Lg_L^{-1})_{G/H}$ and $(\partial_-g_Rg_R^{-1})_{G/H}$. After
some simplification one gets

\eqn\stress{ T_L={1\over 2}tr[((\partial_+t_Lt_L^{-1})_{G/H})^2]=0 ,\qquad
     T_R={1\over 2}tr[((\partial_-t_Rt_R^{-1})_{G/H})^2]=0   }
There remains to parametrize the coset in some convenient way that allows the
solution of these equations.

Up to this point the formalism is general. We now specialize to the ADS model
in $d$ dimensions $SO(d-2,2)/SO(d-1,1)$ and parametrize the coset as in \IBBH\

\eqn\tt { t=\left (\matrix {b  & -bX^\nu \cr
    bX_\mu  & (\eta_\mu^\nu + ab X_\mu X^\nu) \cr } \right ), \qquad
 \partial tt^{-1}=\left (\matrix {0 & -J^\nu \cr
    J_\mu & ab \eta_{\mu\lambda} X^{[\lambda}\partial X^{\nu]} \cr }\right ),
}
where $b=\epsilon (1+X^2)^{-{1\over 2}},\ a=(1-b^{-1})/X^2$, $J^\mu=b(\partial
X^\mu + X^\mu ab X\cdot \partial X) $, $\epsilon =\pm 1$ and
$\eta_{\mu\nu}=diag (1,-1,\dots -1)$ is the Minkowski metric
(when $\epsilon =-1$ the group element is not continuously connected to the
identity; this will play a role in duality). However,
unlike \IBBH\ which uses $X^\mu(\sigma^+,\sigma^-)$, we insert in these
expressions $X_L^\mu(\sigma^+)$ or $X_R^\mu(\sigma^-)$ in order to obtain
$t_L,t_R$. The constraints \stress\ take the form $J_L^2=0=J_R^2$ which
contain non-linear terms and still are difficult to solve. To find more
convenient expressions it is useful to make change of coordinates

\eqn\XLXR{(h_L)^\mu_\nu X_L^\nu b_L =  V_L^\mu, \qquad
    (h_R)^\mu_\nu  X_R^\nu b_R = V_R^\mu   }
where $h_L,h_R$ are Lorentz transformations whose $X_L,X_R$ dependence are
determined through \decoup\ and \tt\ . After using these equations one obtains
the form $h^\mu_\nu J^\nu b=\partial V^\mu $ for both $L,R$ sectors. In this
way the classical constraints simplify to

\eqn\constr{ (\partial_+V_L)^2=0, \qquad   (\partial_-V_R)^2=0 , }
up to overall factors of $b^{-2}$. These are exactly analogous to the
flat string Virasoro constraints and can be solved with identical methods
(e.g. by choosing a light-cone gauge for $V^\mu$). Furthermore, the solutions
\hh\ can be given directly in terms of $V_L,V_R$ by using \decoup\ \tt\ and
\XLXR\ and rewriting

\eqn\hhh{ (\partial h_Lh_L^{-1})^{\mu\nu} = { V_L^{[\mu}\partial
               V_L^{\nu]}\over b_L(1+b_L) }  }
where $b_L=\epsilon (1-V_L^2)^{1\over 2}=\epsilon (1+X_L^2)^{-{1\over 2}}$ and
similarly for the R sector. These equations are solved as path ordered
integrals as in \hh\ . Thus, in terms of $V_L,V_R$ the group elements
$g_L,g_R$ take the form

\eqn\group { g_L=\left (\matrix { b_L & -(V_Lh_L)^\nu \cr
    V_{L\mu}  & [(h_L)_\mu^\nu -  {V_{L\mu} (V_Lh_L)^\nu\over
     1 +b_L }] \cr } \right ) ,   }
and similarly for $g_R$.

It is interesting to point out that this provides a solution to another
non-linear equation. In \IBBH\ we parametrized $g=ht$ with $t$ given in
$(2.8)$. Now $h$ and $t$ depend on both $\sigma^+,\sigma^-$ through
$X^\mu(\sigma^+,\sigma^-)$. It was shown in \IBBH\ that the classical
equations for $X^\mu$ reduce to
\eqn\xeq{ \partial_+ \big [\sqrt {1+X^2} \partial_-\big ({X^\mu\over \sqrt
{1+X^2}} \big ) \big ] = 0 . }
This equation can now be solved for any $d$ by comparing the matrix elements
of $g=ht=g_Lg^{-1}_R$ and reading off $X^\mu$
\eqn\xsol{ X^\mu={1\over b_Lb_R +V_Lh_Lh_R^{-1}V_R }
\big [ (V_Lh_Lh_R^{-1})^\mu - b_L V_R^\mu - {V_Lh_Lh_R^{-1}V_R
\over 1+b_R } V_R^\mu \big ] . }
Using \hhh\ it can be checked that this indeed solves \xeq\ for any $V_L^\mu ,
V_R^\mu $. The expression for $h(\sigma^+,\sigma^-)$ can similarly be read off
by comparing matrix elements of the two parametrizations. It will not be given
here because of its lengthy appearance.

The constraints \constr\ may further be written in the usual Virasoro form
$L_n=0=\tilde L_n$ by introducing normal modes

\eqn\modes{ \partial V_L^\mu = \sum \alpha_n^\mu e^{in\sigma^+}, \qquad
    \partial V_R^\mu = \sum \tilde\alpha_n^\mu e^{in\sigma^-}  }
and writing the usual expressions for $L_n,\tilde L_n$. In a light-cone gauge
one can take $\alpha_n^+=\tilde \alpha_n^+=0$ for $n\ne 0$ and solve for
$\alpha_n^-,\tilde \alpha_n^-$ in terms of the transverse degrees of freedom
and $\alpha_0^+,\tilde \alpha_0^-$ as in the flat string. We have thus
completely solved the {\it classical} ADS strings in terms of
$\alpha_n^\mu,\tilde\alpha_n^\mu$ just like the flat string. The normal modes
introduced here are probably useful in developing a light-cone or covariant
quantum theory as well, but because of the complexity of quantum ordering and
gauge fixing we will not consider this problem in the present paper.


\newsec {Deformed Gauged WZW models and Generalized Duality }

The ungauged WZW action $S_0$ in \naction\ is invariant under
$G_L(\sigma^+)\times G_R(\sigma^-)$ transformations which depend on either
$\sigma^+$ or $\sigma^-$. The gauged WZW action, when $\tilde A_\pm =A_\pm$,
is invariant under the gauge transformations $g'=h^{-1}gh$, $A'_{\pm}=h^{-
1}(A_{\pm}-\partial_{\pm})h$, where the fully local $h(\sigma^+,\sigma^-)$
belongs to the diagonal vector subgroup of $G_L\times G_R$. It is known that
one cannot have full gauge invariance with separate gauge groups for the left
and right transformations. However we could try to embed the subgroup
differently on the left than on the right. Then covariant derivatives of $g$
take the form $D_{\pm}g=\partial_{\pm}g-A_\pm g+g\tilde A_\pm$ where $A_\pm,
\tilde A_\pm$ are related as given in \aa . To generate gauge invariant
equations of motion but with the new covariant derivative we must have the new
action \naction\ with the condition that $Tr(A_+A_-)=Tr(\tilde A_+ \tilde A_-
)$. The new equations of motion are
\eqn\nclass{ (D_+gg^{-1})_{H}=(g^{-1}D_-g)_{\tilde H}=0,\qquad F_{+-}=0,\qquad
D_-(D_+gg^{-1})_{G/H}=0 , }
where one should note the difference in the subscript $\tilde H$ as compared
to \class . This difference means that $Tr(g^{-1}D_-g\tilde t_a)=0$ rather
than $Tr(g^{-1}D_-gt_a)=0$. Furthermore, the covariant derivatives have a new
meaning as pointed out above.

The condition $Tr(A_+A_-)=Tr(\tilde A_+ \tilde A_-)$ together with the
requirement that  $t_a,\tilde t_a$ must have the same commutation rules
place the following constraints on the representation

\eqn\repr{ Tr(t_a[t_b,t_c])=Tr(\tilde t_a[\tilde t_b,\tilde t_c]), \qquad
  Tr(t_at_b)=Tr(\tilde t_a\tilde t_b).  }
Indeed, with these restrictions one finds that the action \naction\ is
invariant under the gauge transformations $g'=h^{-1}g\tilde h$, $A'_{\pm}=h^{-
1}(A_{\pm}-\partial_{\pm})h$ or equivalently $\tilde A'_{\pm}=\tilde h^{-
1}(\tilde A_{\pm}-\partial_{\pm})\tilde h$. Here $h,\tilde h$ contain the same
gauge parameters but they are represented differently. This means that $h^{-
1}\partial_\pm h= j^at_a,\ \tilde h^{-1}\partial_\pm \tilde h= j^a \tilde t_a$
with the same $j^a$, as required to demonstrate the gauge invariance.

What are the solutions to \repr\ apart from the trivial case $\tilde t_a=t_a$
? It is immediately clear that for a $U(1)$ subgroup one can take $\tilde t=\pm
t$ in accordance with the well known vector or axial gauging. Therefore, we
concetrate on the non-Abelian part of the gauge group. One
solution is $\tilde t_a=g^{-1}_0t_ag_0$ where $g_0$ is any global group
element (independent of $\sigma^+,\sigma^-$) in {\it complexified} $G_R$.
Furthermore $g_0$ can also include discrete transformations that are not
connected to the identity.  In addition, note that the transpose
representation will be a new candidate for $\tilde t_a=-(t_a)^T$, which can
then be conjugated by any element $g_0$. Thus, we have the solutions

\eqn\dual{ \tilde t_a = g^{-1}_0t_ag_0 , \qquad  \tilde t_a =- g^{-
1}_0(t_a)^Tg_0  . }
If there are any additional automorphisms of specific Lie algebras that are not
obtainable by \dual\ they will provide additional solutions for
$\tilde t_a$ (which may also be conjugated by some $g_0$). We recognize
that \dual\ generalizes the vector/axial options that were available for
$U(1)$.

Some of these options are not independent. For example consider the $2d$
black hole, with the $SO(2,1)$ generators $J_0,J_1,J_2$, where $J_0$ is
antisymmetric and $J_1,J_2$ are symmetric real $3\times 3$ matrices, and the
non-compact $J_2$ gauged. The passage from vector to axial gauging may be
achieved by either a $g_0$ corresponding to a $180^o$ rotation around
$J_0$ ($\tilde J_2= -J_2$) or by the second choice in \dual\ corresponding to
transposition and $g_0=1$. The passage from axial to vector may also be done
continuously with a $g_0=exp(\alpha J_0)$. Then $\tilde J_2= cos\alpha
J_2+sin\alpha J_1$. With this choice, taking a unitary gauge for $g$
as before [6,9,10,18,19], and inserting it in the new action \naction\ one
finds the metric and dilaton

\eqn\metr{ \eqalign {&ds^2=dr^2-dt^2 {cos\alpha\ cosh(2r) +1\over cos\alpha\
cosh(2r) - 1}  + dtdr {2 sin\alpha \over cos\alpha\ cosh(2r) -1 }\cr
   &\Phi = ln\big (cos\alpha\ cosh(2r) - 1\big ) }  }
that shows the role of a continuous ($\sigma^+,\sigma^-$ independent)
parameter in generating deformed metrics. In particular for $\alpha =0,\pi$ we
get the previous vector/axial case respectively.  For general $\alpha$ the
deformed metric \metr\ does not represent a new manifold, as can be seen by
finding the Kruskal coordinates and noting that it takes the standard form
given by Witten \WIT .

The role played by $\alpha =0$ (original action) and $\alpha =\pi$ (dual
action) is reproduced by the sign $\epsilon =\pm 1$ of the previous section
(see the expression for $b$ following \tt\ or \hhh\ ) in the unitary gauge.
Namely, if the group element is connected to the identity continuously
($\epsilon=1$) one obtains the metric with $\alpha=0$ and if the group element
is in a class disconnected from the identity ($\epsilon=-1$) one gets the
metric with $\alpha=\pi$. Therefore, the original action describes both
metrics provided all group elements are included (as already known for the
2d black hole in Kruskal coordinates \WIT ). On the other hand, if one
insists on only group elements connected to the identity, then the dual
actions taken together reproduce the full manifold as parametrized by Kruskal
coordinates.

Let us explore a bit further the role of $g_0$ as deformation parameters
versus a duality transformation. Using the invariance of $S_0$ under
$G_L\times G_R$ one may rewrite the full action \naction\ as follows (with
$\tilde A_+=g^{-1}_0A_+g_0$)

\eqn\nnaction{ S(g,A^a_\pm,g_0)= S(gg^{-1}_0, A^a_\pm,1) . }
This is {\it not} a statement of symmetry (since the result is not
$S(g,A^a_\pm,g_0=1)$). In particular if $g$ is a solution of the $g_0=1$
action  then $gg^{-1}_0$ is not a solution of the same action. Instead, it is
a solution of another action that differs from the original one by $g_0$.
However, we may expect that there are some $g_0$ for which we get dual
theories. As a criterion for duality we require that the space of classical
solutions (or normal modes) of the two dual theories be identical (as for the
axial/vector case of the 2d black hole \IBBH\ ). Thus we look for special
$g_0$ for which the classical solution space is the same in the presence or
absence of $g_0$. Solving  equations \nclass\ partially $F_{+-}=0$ and
choosing the gauge that gives $A_\pm=0$ (see the previous section) one finds
that the only difference in the remaining equations is
\eqn\duality{ Tr(g^{-1}\partial_-gg_0^{-1}t_ag_0)=0 \quad {\rm versus}\quad
             Tr(g^{-1}\partial_-gt_a)=0 . }
Thus the equations span the same set only when $g_0^{-1}t_ag_0$ is some linear
combination of the $t_a$ and is orthogonal to any generator in $G/H$. For
such $g_0$, when they are not trivial global rotations embedded in the gauge
subgroup $H$, one obtain a duality transformation. One such example is the
$\alpha=\pi$ rotation illustrated above.

This classical duality argument applies also to the case involving the
transpose representation. In this case the requirement analogous to \duality\
is automatically satisfied since, in an appropriate basis, $-(t_a)^T$ differs
from $t_a$ at most by some minus signs. Therefore for this case we have
automatically dual theories. The same set of $g_0$'s as above may also
generate additional dual theories. In some instances it is possible to prove
that $-(t_a)^T=g_0^{-1}t_ag_0$ for some $g_0$. This is certainly the case for
ADS strings with e.g. $g_0=diag(\pm 1,-1,1,1,\cdots, 1)$ where either sign
$\pm$ in the first entry will do the job. However, this is not always the
case, for example for SU(N) $-(t_a)^T$ is equivalent to the complex conjugate
representation which may not be obtained via a similarity transformation.

The path integral based on the dual theory with $-(t_a)^T$ is not obviously
identical to the original one. However, on the basis of the above classical
argument, we expect that by rewriting the path integral as in \KIR\ that the
duality can be shown. As mentioned in the previous paragraph in some cases $-
(t_a)^T$ is obtained with a similarity transformation. Let us consider such
cases along with the arbitrary $g_0$ deformations. In a quantum theory defined
by a path integral one integrates over all group elements, not only solutions.
Then at first sight there seems to be no difference between $g$ and $gg_0^{-
1}$ and therefore one might expect that the new action and the old one give
equivalent quantum theories. This would have been the  simplest proof of
quantum duality for the ADS strings (and in particular of the 2d black hole)
if the integration measure were the Haar measure for $g$. However there is no
reason to assume that the measure be invariant under left or right
multiplication with arbitrary group elements. After all, since the action is
invariant only under the diagonal (or deformed diagonal) subgroup $H$, it is
sufficient to require an invariance of the measure only under $H$. In fact the
criterion for the correct measure is conformal invariance including the
effects of the gauge field measure. As we shall argue in the next section,
indeed the correct measure is not the Haar measure. Therefore, to prove
duality at the quantum level one needs to rewrite the path integral as in the
example provided in \KIR . We will postpone this proof to future work.


\newsec {ADS Strings in 3 dimensions}

We now study the ADS string in 3 dimensions based on
$SO(2,2)_{k}/SO(2,1)_k$. As pointed out in \BN\ there is a generalization of
this model with two different central extensions in the form
$SO(2,1)_{k_1}\times SO(2,1)_{k_2}/ SO(2,1)_{k_1+k_2}$. The $k_1\ne k_2$
version was recently examined
\ref\crescimano{ Crescimano, ``Geometry and Duality of a Non-Abelian Coset
Model", LBL-30947.}.
We will analyse the $k_1=k_2$ version in a parametrization that leads to a
diagonal metric. Furthermore, we will discuss the duality properties and the
new metrics that follow from the new action \naction .

At first we take $\tilde A_\pm =A_\pm$. We parametrize $g=ht$ with $t$ given
in $(2.8)$. Furthermore, for $3d$ the Lorentz transformation $h$ may be
constructed from a vector
\def\e'{{\epsilon}'}
\eqn\lorentz { h_{\mu\nu}= \e'\sqrt {1-Y^2} \eta_{\mu\nu} + {Y_\mu Y_\nu \over
1+\e'\sqrt {1-Y^2} } + \epsilon_{\mu\nu\lambda} Y^\lambda , }
where $\e'=\pm 1$ as in \tt\ (signifying group elements connected or
disconnected from the identity). In terms of this parametrization the Haar
measure for $SO(2,2)$ turns out to be
\eqn\gmeasure{ (Dg)= {(D^3X)(D^3Y)\over |f(X,Y)|},\qquad
                  f= (1+X^2)^2\sqrt {1-Y^2} (1+\e'\sqrt {1-Y^2})         }
where $(D^3X),(D^3Y)$ are the naive Minkowski or Euclidean measure. As pointed
out at the end of the last section, one is not required to take the Haar
measure since the action is invariant only under the $H$ subgroup. Therefore,
one may modify this measure by any Lorentz invariant factor $F(X^2,Y^2,X\cdot
Y)$. As will be seen later, this will be necessary because of conformal
invariance which will fix $F$. The action contains a quadratic term of the
form $A_+^aM_{ab}A_-^b$. Integrating out the $A_\pm$ gives a determinant that
modifies the measure in the path integral by a factor
\eqn\dilaton { det(M)^{-1} = { (1+X^2)(1+\e'\sqrt {1-Y^2})\over (X^2 Y^2 -
(X\cdot Y)^2)}. }
As we shall see, this gauge invariant expression will be identified with the
dilaton field $Ce^{-\Phi}=det(M)^{-1}$, where $C$ is a constant. To obtain the
effective action one solves the first pair of equations in \class\ and
substitutes the solution for $A_\pm$ in the action. The gauge part of the
action then reduces to
 \eqn\gaugeaction{ S_1=-{k\over 4 \pi} \int Tr(A_-\partial_+gg^{-1}) }
where $A_-$ is the solution. It was shown in \IBBH\ that the $A_-$ equation
becomes
\eqn\adseq { (h^{-1}D_-h)^{\mu\nu}=a X^{[\mu}(D_-X)^{\nu]} }
where $(D_\pm X)^\mu = \partial_\pm X^\mu - (A_\pm)_\nu^\mu X^\nu$ and $(D_\pm
h)_\mu^\nu = \partial_\pm h_\mu^\nu - [A_\pm,h]_\mu^\nu $. This equation can
be solved for $A_-$ without fixing any gauges. Substitution in \gaugeaction\
leads to a non-linear locally Lorentz invariant action written in terms of the
two vectors $X,Y$. We have done this, but because of the complexity of the
expression we specialize to a gauge fixed form.

Local Lorentz transformations cannot change the sign or magnitude of
$X^2,Y^2$. Therefore the cases of $X^2>0,\ X^2<0,\ X^2=0$ (similarly for
$Y^2$) have to be gauge fixed separately. Fortunately, it is possible to pass
from one sign to the other by analytic continuation of the gauge fixed
parameters. Therefore, we concentrate on space-like $X^2,Y^2< 0$. By Lorentz
transformations these two vectors can be brought to the form
\eqn\fix{ X^\mu = tanh(2r)\ (0,cos\theta , sin\theta), \qquad Y^\mu = \e'
 sinh(2t) \ (0 , 0 , 1 ). }
In choosing this form we were inspired by the two dimensional black-hole
(specializing to $\theta =0$ yields the two dimensional case with $r,t$
identified with the parameters used in two dimensions). After some tedius
algebra we obtain the gauge fixed form of the action \naction
\eqn\action{ S={k\over \pi} \int d^2\sigma  \ \
G_{\alpha\beta}\partial_+\phi^\alpha\partial_-\phi^\beta , \qquad
\phi^\alpha=(r,\theta,t),  }
where the target space metric $G_{\alpha\beta}$ is given by the line element
$ds^2=G_{\alpha\beta}d\phi^\alpha d\phi^\beta$ and the antisymmetric tensor
$B_{\alpha\beta}$ is zero with our parametrization. However, the form that
emerges for $G_{\alpha\beta}$ depends on the signs $\epsilon , \e'$ chosen in
\tt\ and \lorentz\ as follows
\eqn\metric{ \eqalign { & ds^2=dr^2+\lambda^2(r,\epsilon ) \ [d\theta +
\kappa (t,\e')\ tan\theta\ dt]^2 - { 1\over \lambda^2(r,\epsilon )\
 cos^2\theta} dt^2 \cr
& \lambda^2(r,\epsilon )= {cosh(2r)- \epsilon \over cosh(2r)+ \epsilon} \qquad
 \kappa (t,\e')= {sinh(2t)\over cosh(2t)-\e'} } }
Depending on these signs the various factors simplify to
\eqn\signs{ \lambda^2(r,1) = tanh^2r, \quad
     \lambda^2(r,-1) = coth^2r ,\quad
     \kappa (t,1) = coth(t), \quad  \kappa (t,-1) = tanh(t). }
These forms clearly indicate that the various signs $\epsilon ,\e'$ are
closely related to duality transformations. In fact, we find that the effect
of $\e'$ is reproduced by a dual theory defined with a $g_0=(1,-1,-1,1)$
as in section 3. The effect of $\epsilon$ is also reproduced by a dual theory
with $g_0=(-1,1,1,-1)$ provided one works in the gauge $X^\mu = tanh(2r)\
(0,0 ,1), Y^\mu = \e'sinh(2t)\ (0,-sin\theta,cos\theta)$ instead of \fix .
Both of these gauge choices lead to the same metric \metric\ in the original
theory.

The metric can be put into diagonal form by changing variables
\eqn\change{ T=\sqrt{cosh(2t)+\e'\over 2},\qquad z=\sqrt{cosh(2t)-\e'\over
2}sin\theta }
which gives
\eqn\diag{ ds^2=dr^2+ {\lambda^2(r,\epsilon)\ dz^2 - \lambda^{-2}(r,\epsilon)\
dT^2 \over T^2-z^2-\e'}  }
where $T^2-z^2-\e' = {1\over 2}(cosh(2t)-\e')\ cos^2\theta\ \ge 0$.

The dilaton (which should come out from the integration measure) must satisfy
the target space Einstein equation $R_{\alpha\beta}=D_\alpha D_\beta \Phi$
perturbatively in powers of $1/k$ in order to satisfy conformal invariance
 \ref\callan{ C.G. Callan, D. Friedan, E.J. Martinec and M. Perry, Nucl. Phys.
B262 (1985) 593.}.
Obviously, this equation is simpler in the new basis $\phi^\alpha=(r,z,T)$.
In the two dimensional black hole $\Phi=ln(detM)+ const. $ solves the dilaton
equation. Taking this as a guide we try the form \dilaton\ after using \fix\
and \change
\eqn\dil{ e^{\Phi-\Phi_0} =(detM)= sinh^2(2r) \ cos^2\theta \ (cosh(2t) - \e')
= 2 \ sinh^2(2r)\  (T^2-z^2-\e')  }
After tedius algebra one finds that indeed \dil\ solves the curvature equation.
Also the central charge of the model can be computed (to leading order in
powers of $1/k$) using \callan\
\eqn\central{ c=3+{3\over {2\ k}}(D_{\alpha}\Phi D^{\alpha}\Phi
+D_{\alpha}D^{\alpha}\Phi).}
We find $c=3+{18\over k}$ in agreement with the expansion of the exact central
charge $c={6k\over {k-4}}-{3k\over {k-2}}$ in powers of $1/k$. Next we compute
the scalar curvature
\eqn\curvature{ R= -{8\over sinh^2(2r)}\big [1+{2\epsilon cosh(2r)\ (T^2+z^2)-
\e' (cosh^2(2r)+1)\over 2(T^2-z^2-\e') } \big ] . }
One sees that there are singularities at $r=0$ and at $T^2-z^2-\e'=0$
(equivalently at $\theta=\pm {\pi\over 2},\ t=0 $ for $\e'=1$, or $\theta=\pm
{\pi\over 2}$ for $\e'=-1$  ). We believe that this is a new singular solution
of Einstein's equations
\eqn\einstein{ R_{\alpha\beta}-{1\over 2}G_{\alpha\beta} R = T_{\alpha\beta} .}
The matter energy momentum tensor in this equation does not vanish. Therefore,
unlike the two dimensional black hole which corresponds to a vacuum solution,
in the present case there is non-trivial matter.

As mentioned earlier, by chosing the gauge \fix\ we have so far concetrated on
the space-like regions $-1< X^2<0,\ Y^2<0$. Note that $1+X^2>0$ and $1-Y^2>0$
must always be satisfied for well defined $SO(2,2)$ group elements in \tt\ and
\lorentz . The remaining regions can be obtained either by choosing appropriate
gauges or by simply doing analytic continuation in the variables $r,t,\theta$
while making sure that one remains within $SO(2,2)$ (in a transformed basis).
Thus, for $-1<X^2<0,\ 0< Y^2< 1$ substitute $(r,t,\theta)\rightarrow
(r,i\theta, it)$; for $0<X^2,\ 0<Y^2<1$  take $(r,t,\theta)\rightarrow (it,
i\theta,ir)$ and finally for $0<X^2,\ Y^2<0$ replace $(r,t,\theta)\rightarrow
(it,r,i\psi)$ where $\psi$ is non-compact. The metrics obtained from \metric\
through these substitutions all have a single time coordinate designated by
$t$. However, due to analytic continuation, not every patch has the same
structure of curvature singularities. It is also understood that
the metrics for coset theories based on the groups $SO(3,1)$ and $SO(4)$
follow by appropriate analytic continuations of our expressions. Curvature
singularities are seen to occur also for these cases.

The analytically continued metrics together with the duality transformed
metrics give various patches of the complete space-time manifold (as for the
$2d$ black hole). The shape of the manifold is clearly complicated. Consider
the patch given by \metric\ . At constant time $dt=0$ it describes the shape
of a semi-infinite cigar for $\epsilon=1$ or a trumpet for $\epsilon=-1$ (like
the 2d Euclidean black hole metric). However, this shape does not remain
static and it evolves with time. It rotates at a rate which is both $t$ and
$\theta$ dependent. The meaning of the singularities are presently unclear. To
further interpret this manifold one needs to find its global properties by
studying the geodesics of light rays, finding the geodesically complete metric
and writing it in terms of global coordinates (analog of Kruskal coordinates).
This remains for future work.

We now turn to an analysis of the measure. The gauge fixing $(X^0,Y^0,Y^1=0)$
introduces a Faddeev-Popov determinant in the path integral measure which may
be computed from the Lorentz transformation properties
\eqn\fp{ \left (\matrix {\delta Y^0\cr \delta Y^1\cr \delta X^0}\right )=\left
(\matrix {0 & Y_2 & 0 \cr -Y_2 & 0 & 0 \cr 0 & X_2 & -X_1}\right ) \left
(\matrix {\omega_{12} \cr \omega_{02} \cr \omega_{10} } \right ) }
where $\omega_{\mu\nu}$ are the infinitesimal Lorentz parameters. Thus, the
Faddeev-Popov determinant is
\eqn\fpdet{ \Delta = X^1(Y_2)^2 . }
The remaining integrations are $(DX^1)(DX^2)(DY^2)=(Dr)(Dt)(D\theta)\ J$,
where $J$ is the Jacobian due to the change of variables
$ (X^1,X^2,Y^2)\rightarrow (r,\theta,t)$ given in \fix\
\eqn\jacob{ J=|{sinh(2r)\ cosh(2t)\over cosh^3(2r) }| }
Putting all the factors together the integration measure takes the form
\eqn\measure{ (Dr)(Dt)(D\theta)\ J\ \Delta \ det(M)^{-1} \ f^{-1}\ F }
where all the factors, except F, are computed above. It is interesting to
observe that the known factors combine to give $D^3\phi \sqrt {-G} F$
\eqn\detg{ J\Delta (detM)^{-1}f^{-1}=\sqrt {-G} = {1\over |cos\theta |} .}
A similar relation holds for the two dimensional black hole as well \KIR\ , so
maybe one can prove a theorem for any gauged WZW model. Assuming that
this relation is general, we give a nice prescription for computing the
dilaton without directly evaluating the determinant $detM$. Namely,
\eqn\dila{ e^{-\Phi}=(detM)^{-1}=\sqrt {-G} J^{-1}\Delta^{-1} f. }
This can serve either as a calculational tool or as a check.
Finally we are in a position to determine F. Since the dilaton must be
identified by exponentiating the entire measure $\sqrt{-G} F$, one
must choose
\eqn\fff{ F = J^{-1}\Delta^{-1} f = {  e^{-\Phi} \over  \sqrt {-G} }.  }
For consistency with Lorentz invariance of the original measure this must be a
function of only dot products involving $X^\mu,Y^\mu$. This is indeed the case
since $exp(-\Phi)$ is written in terms of dot products in \dilaton\ and $(-
G)^{-{1\over 2}}=cos\theta =[1-(X\cdot Y)^2/X^2Y^2]^{1\over 2}$. A non-trivial
$F$ is necessary for the $2d$ black hole as well. This factor in the measure
seems to have been neglected in previous path integral discussions of gauged
WZW models.

We hope to extend these results to the four dimensional case and to
investigate the quantum theory.

\newsec {Note Added}

At the time of preparation of our paper we received a manuscript by Fradkin
and Linetsky \ref\fradkin{E.S. Fradkin and V.Ya. Linetsky, ``On Space-Time
Interpretation of the Coset Models in $D<26$ String Theory", HUTP-91/A044}
which overlaps with some of our results. In particular there is agreement
with some patches of the three dimensional metric and dilaton even though the
parametrization of the group element and the details of the calculations are
quite different.

\listrefs
\end